\documentclass[acmsmall, anonymous=false, authorversion=false]{acmart}

\AtBeginDocument{%
  \providecommand\BibTeX{{%
    \normalfont B\kern-0.5em{\scshape i\kern-0.25em b}\kern-0.8em\TeX}}}

\copyrightyear{2021}
\acmYear{2021}
\acmDOI{TBD}

\acmConference{CSCW '21 (Pre-print)}
\acmVolume{}
\acmNumber{}
\acmArticle{}
\acmMonth{}

\begin{document}

\title[Beyond the Command]{Beyond the Command: Feminist STS Research and Critical Issues for the Design of Social Machines}

\author{Kelly B. Wagman}
\email{kbwagman@mit.edu}
\orcid{}
\affiliation{%
  \institution{Massachusetts Institute of Technology}
  \streetaddress{77 Massachusetts Ave}
  \city{Cambridge}
  \state{MA}
  \postcode{02139}
}

\author{Lisa Parks}
\email{lparks@mit.edu}
\orcid{}
\affiliation{%
  \institution{Massachusetts Institute of Technology}
  \streetaddress{77 Massachusetts Ave}
  \city{Cambridge}
  \state{MA}
  \postcode{02139}
}


\begin{abstract}
Machines, from artificially intelligent digital assistants to embodied robots, are becoming more pervasive in everyday life. Drawing on feminist science and technology studies (STS) perspectives, we demonstrate how machine designers are not just crafting neutral objects, but relationships between machines and humans that are entangled in human social issues such as gender and power dynamics. Thus, in order to create a more ethical and just future, the dominant assumptions currently underpinning the design of these human-machine relations must be challenged and reoriented toward relations of justice and inclusivity. This paper contributes the "social machine" as a model for technology designers who seek to recognize the importance, diversity and complexity of the social in their work, \textit{and} to engage with the agential power of machines. In our model, the social machine is imagined as a potentially equitable relationship partner that has agency and as an "other" that is distinct from, yet related to, humans, objects, and animals. We critically examine and contrast our model with tendencies in robotics that consider robots as tools, human companions, animals or creatures, and/or slaves. In doing so, we demonstrate ingrained dominant assumptions about human-machine relations and reveal the challenges of radical thinking in the social machine design space. Finally, we present two design challenges based on non-anthropomorphic figuration and mutuality, and call for experimentation, unlearning dominant tendencies, and reimagining of sociotechnical futures.
\end{abstract}

\ccsdesc[500]{Human-centered computing~HCI theory, concepts and models}

\keywords{feminism, feminist HCI, science and technology studies, human-robot interaction, design}

\maketitle

\section{Introduction}
"Alexa, tell me the weather!" has become a common command. By January 2019 over 100 million devices equipped with Amazon’s virtual digital assistant, Alexa, had been sold worldwide \cite{Bohn_2019}. While seemingly simple, this human-machine interaction, in which a human voice orders an artificially intelligent digital assistant to instantly deliver information, is deceptively complex. Alexa’s human-like voice is gendered feminine and she performs historically feminized clerical labor. This interaction both depends on and impacts global material conditions: the Alexa device hardware is sourced from numerous countries; the software relies on layers of physical internet infrastructure and value-laden machine learning algorithms; and the discarded devices and data centers pump toxins into the environment \cite{Crawford_Joler_2018}. Beyond this, the interaction raises a fundamental question: what exactly is Alexa’s relationship to the user and to humans, more generally? Our social norms do not yet include clear conventions for how to interact with digital assistants or robots, or what we call in this paper, "social machines." It is not even immediately obvious to most people that the command "Alexa, tell me the weather!" may be problematic. Yet, as we suggest, this brief example evokes a host of critical issues related to gendered and other social power dynamics in human-machine relations.

For decades feminist scholars have critiqued science and technology, yet technology design outputs have been largely unresponsive to these critiques. Feminists in scientific and technology fields have called for gender diversity in the workforce (\textit{e.g.,} gender balanced design teams); gender diversity in the "substance of science" (\textit{e.g.,} a digital assistant that helps with questions about reproductive health); and feminist approaches to methods and design practices (\textit{e.g.,} not universalizing "the user" in design methods) \cite{Schiebinger_1999}. While each area is important, our contribution in this paper falls within the third area because it questions how foundational assumptions about human-machine relations and structural conditions impact technology design. Some key foundational assumptions that we challenge include: that machines are politically neutral; that machines cannot form social relationships; that machines do not have agency; that humans should control machines; and that there is a clear boundary between human and machine. Drawing on feminist STS scholarship (\textit{e.g.,} \cite{Balsamo_1996, Benjamin_2019, Browne_2015, Schwartz-Cowan_1983, Wajcman_2004, Haraway_1991}), we explore how power works in human-machine relations and suggest that fuller awareness of the social can enhance technology design. We argue that HCI scholars and others concerned with technology design must confront the fact that common assumptions about the role of machines in the world reinforce existing inequalities, injustices, and patterns of oppression. Because of this, we must consider radical shifts in our thinking and approaches to design, and set out to craft machines, and engage in human-machine relations, in more ethical, just, and inclusive ways \cite{Buolamwini_2018, Costanza-Chock_2020}. 

Our main contribution in this paper is a conceptual model for human-machine relations that operationalizes key lessons from feminist STS in ways that are generative for designers and technology builders.  We consider designers to be crafting "social machines" and "human-machine relations" as opposed to simply building "machines." All machines are part of the \textit{social}, a science and technology studies (STS) term that broadly refers to what is produced when humans and non-humans interact and develop relationships, and become part of power relations, societal norms, and cultures \cite{Latour_2005}. In this framework, interacting humans and non-humans are \textit{mutually shaping}; humans and non-humans both influence, and are influenced by, one another. In the "Alexa, tell me the weather!" example, the conversation between Alexa and a user is a social interaction, as are the relations between Alexa’s designers and developers at Amazon and factory workers producing the devices at Foxconn \cite{Togoh_2019}. Throughout the device’s lifecycle, Alexa can alter the lives of humans \textit{and} the norms and practices of those humans in turn inform Alexa’s development.

To account for these conditions, we propose use of the term "social machine" as an actionable design intervention. By using this term, we make "the social" explicit and encourage  technology builders to rigorously reflect upon and engage with relations of mutuality in their work. We define "social machine" as an object that is designed to construct and engage in social relations with humans, and that has been crafted with careful attention to issues of agency, equitability, inclusion, and mutuality.\footnote{Note that we are using \textit{social machines} differently than Judith Donath in her book \textit{The Social Machine: Designs for Living Online} \cite{Donath_2014}. She is referring to machines that function as a communication medium and allow for social interaction between humans, while we center on a machine that is itself social. We also do not mean collections of people and machines that function together to create a social machine, as in \cite{Shadbolt_Smith_Simperl_VanKleek_Yang_Hall_2013}.} The term "social machine" is also meant to recognize the proliferating human-machine relations that take shape in the digital era via computing interfaces, artificial intelligence, digital assistants, and robotics, but it is not meant to be an essentialist term. There is no physical characteristic that makes one object a social machine and not another; rather, any object can be a social machine if it is designed with consciousness of social inequalities and injustices, and partakes in a purposeful effort to remedy them. For example, the Amazon Alexa device as it stands is \textit{not} a social machine since it was not designed with equity and inclusion in mind; however, this could change if Alexa were re-designed with greater emphasis on social differences and power dynamics. The material sites of human-machine relations construct, operationalize, analyze/iterate, and naturalize/normalize different kinds of relations, some of which reproduce oppression.

There are existing terms that are related to the idea of a social machine, but they tend to essentialize the human-machine binary. They include social/sociable robot \cite{Breazeal_2002} and relational artifact \cite{Turkle_Taggart_Kidd_Daste_2006}. We use "machine" as opposed to "robot" in order to avoid assumptions about robots that are baked into the term’s history; namely, that they serve humans by performing mechanized labor (Oxford English Dictionary); that they are embodied and anthropomorphized \cite{Breazeal_2002}; and that they are fundamentally different from earlier machines like the computer or sewing machine and thus should be studied separately. Turkle’s notion of a "relational artifact" evocatively suggests that various kinds of machines can present themselves as having "states of mind" and insists that human awareness of this possibility can enrich their encounters with machines. Turkle’s work on "relational artifacts" is influenced by developmental psychology and psychoanalysis and ultimately is concerned with the ways that \textit{humans} benefit from or are harmed by these encounters.\footnote{Turkle’s notion of relational artifacts refers to "artifacts that present themselves as having `states of mind' for which an understanding of those states enriches human encounters with them" \cite{Turkle_Taggart_Kidd_Daste_2006}. The term is intended to highlight "the psychoanalytic tradition, with its focus on the human meaning of the artifact–person connection" \cite{Turkle_Taggart_Kidd_Daste_2006}.} Our model builds on work in feminist STS\footnote{In STS "sociotechnical relations" \cite{Bowker_Star_2000} refers to the ways social forces and technological objects, systems, and practices dynamically shape and inflect one another. Our definition of social machines builds from this idea, but is more specific and is meant as an intervention in contemporary design practices.} that conceptualizes technical artifacts as deeply embedded within their social contexts and thus within relations of power. Again, we use the term "social machine" to underscore the necessity of engaging with the social and thus issues of equitability, mutuality, and inclusion in the design process. 

Our model posits a social machine as a non-human "other" that is distinct from, yet related to, humans, objects, and animals. The social machine has agency to act in the world and is conceptualized as having an \textit{equitable} potential and inclusive position alongside humans and other non-humans. Our model stands in contrast to existing models of human-machine relations that conceptualize the machine as a tool, as a human companion, as an animal or creature, or as a slave. Existing models are problematic, we argue, because they either imply a domination of the human over the machine, fail to recognize the machine as distinct from humans/animals, or do not acknowledge machine agency. Grounded in feminist STS perspectives, our model is not merely a critique of an existing system, but, with its emphasis on design, offers a generative way to think about new forms social machines could take, based on an ethics of inclusivity, equitability, and mutuality. 

To aid designers in building and supporting new kinds of human-machine and social relationships, we seek to bridge feminist science and technology studies (STS) research with computer-supported cooperative work (CSCW), human-computer interaction (HCI), and human-robot interaction (HRI). The fields of HCI, CSCW,\footnote{EunJeong Cheon and Norman Makoto Su \cite{Cheon_Su_2017} explored how roboticists try to understand the imagined users of their robots, and how this process in turn shapes robot designs. Martin Porcheron, Joel Fischer, and Sarah Sharples \cite{Porcheron_Fischer_Sharples_2017} have studied how digital assistants function when part of a conversation occurring among several humans. Other researchers have examined how robots integrate into workplace teams as surgical robots, collaborative automation, and nurse assistants \cite{Cheatle_Pelikan_Jung_Jackson_2019, Mackeprang_Muller-Birn_Stauss_2019, Taylor_Lee_Kubota_Riek_2019}.} and HRI,\footnote{HRI’s principle framework assumes that robots are entirely distinct from humans \cite{Bartneck_Belpaeme_Eyssel_Kanda_Keijsers_Sabanovic_2020}, limiting design potential. HRI also exhibits an implicit embrace of technological determinism \cite{Sabanovic_2010} by presuming that a robot will inevitably affect its surroundings, yet there is not equal acknowledgement of the manner in which the robot’s design has been shaped by the researchers’ own social norms and biases. Significantly, some HRI research pushes back against these assumptions (\textit{e.g.,} \cite{Forlizzi_DiSalvo_2006, Sabanovic_2010, Sabanovic_Bennett_Lee_2014, Allouch_deGraaf_Sabanovic_2020}), but it is infrequent and tends to sidestep the power relations of the social.} have variously begun to confront the social dimensions of machines. We see an opportunity to unite these fields together with feminist STS, especially when addressing the design of social machines. A decade ago Shaowen Bardzell \cite{Bardzell_2010} put forth a feminist HCI research agenda that delineated a series of generative design principles to improve design methods from a feminist perspective. Since then, feminist HCI has been extended to encompass humanistic and emancipatory HCI \cite{Bardzell_Bardzell_2015, Bardzell_Bardzell_2016}, which advocate for anti-oppressive technology and address other axes of social difference beyond gender, including race, class, sexuality, and ability, among others. Other scholars, too--in CSCW, HCI, and design studies--have explicitly called their work \textit{feminist} or \textit{emancipatory}, revealing potential for more critical awareness and transformative design work \cite{Costanza-Chock_2020, Irani_Vertesi_Dourish_Philip_Grinter_2010, Keyes_Hoy_Drouhard_2019, Kumar_Karusala_Ismail_Wong-Villacres_Vishwanath_2019, Rosner_2018, Schlesinger_Edwards_Grinter_2017}. Much of this work has centered on reflexive design practices that undo harmful dominant assumptions; our paper continues in this tradition, but specifically delineates what a human-machine relation is/can be in a design context and offers the "social machine" as an alternative to dominant models. 

In what follows we apply critical interpretive methods to review and evaluate the treatment of social issues and feminist design possibilities across several scholarly fields. These methods involved reviewing scholarly literature across CSCW, HCI, HRI, and feminist STS, isolating research and concepts relevant to the design of social machines, questioning foundational assumptions in these fields, and using this process to formulate a theoretical model and design challenges. We begin with an overview of feminist science and technology studies (STS) scholarship on human-machine relations. We then discuss the implications of this scholarship for technology design and HCI research. We argue that feminist STS research offers designers/HCI scholars ways of thinking about the complexities of the social, user identities, and power dynamics that can enrich the process of conceptualizing, designing, and developing social machines. In the next section, we use a feminist STS lens to examine and critique four dominant categories of human-machine relations in robotics, including machine as tool, as human companion, as animal or creature, and as slave. Following this discussion, we articulate our own model, positing the social machine as an "other" to humans, objects, and animals; an actor with agency; and as a potential equal in power relations with humans. Finally, we propose concrete design challenges involving non-anthropomorphic figuration and relations of mutuality in order to inspire future work experimenting with our model.

\section{Background: An Overview of Feminist STS Research on Human-Machine Relations}

Our model is grounded in a tradition of feminist STS research on human-machine relations that began during the 1980’s. Since our model aims to provide a blueprint for more ethical and just human-machine relations, we begin by recognizing key insights made by feminist STS scholars who have been studying the gendering of technology for decades. Our goal here is not to provide a comprehensive historical analysis of gender, technology, and automation, rather, we want to emphasize feminist STS scholarship that has inspired our model. In this section, we draw this scholarship to: demonstrate how the history of gendered labor and social inequalities resulted in technologies that disenfranchised women; explore how cultural binaries (like male/female) have been critiqued and replaced; and emphasize the importance of intersectional feminisms, which demand inclusion of race/ethnicity in understandings of human-machine relations. We draw on feminist STS scholarship to elaborate an actionable conceptual model for technology designers/builders.

\subsection{From the gendered division of labor to sociotechnical relations}

Early feminist STS research focused on histories of the gendered division of labor and explored how men and women have been positioned differently in relation to various technologies. In modern Western industrial societies, men historically held jobs in the public sphere, whether in governance, finance, or on assembly lines, and women typically performed labor invisibly in the private sphere of the home. Though women entered the public sector workforce in greater numbers during the late 19th and early 20th centuries, the professional technology workforce and fields such as computer engineering, software development, and user-interface design continue to be dominated by men \cite{Hicks_2017}. While women have always worked, for centuries childcare and domestic labor were unrecognized as “work” and were generally unpaid. Women’s domestic labor was not counted in formal economic measures such as GDP, but, as feminist STS scholars have shown, women have always been involved with machines, whether looms and sewing machines, typewriters, or television sets \cite{Schwartz-Cowan_1983, Spigel_1992, Wajcman_2010}. More recent research has explored women’s crucial yet under-recognized roles in the history of computing \cite{Hicks_2017, Nooney_2020, Shetterly_2016}.

In an effort to complicate reductive notions of technology as a tool, feminist scholars have worked to deepen understandings of technology’s relation to the social. Extending work by systems historians and theorists \cite{Hughes_1993} and social constructionists \cite{Bijker_Hughes_Pinch_1987, Pinch_Bijker_1984}, feminists have approached technology as an artifact and practice that is both embedded within and has the potential to shape social relations \cite{MacKenzie_Wajcman_1985}. Feminist STS scholars also have been influenced by Bruno Latour’s "actor-network theory," which understands “technology” within a network of relations involving human and non-human actors \cite{Latour_2005}. Technology, thus, went from being considered a stable technical object to a dynamic web of interrelations involving organizations, finance, labor, cultural norms, and artifacts, like the global Amazon Alexa ecosystem described in the introduction (see \cite{Crawford_Joler_2018} for an example). This web of interrelations became known as a sociotechnical system (\textit{e.g.,} \cite{Bowker_Star_2000, Wajcman_2010}). One of the significant moves in feminist STS work has been to insist that technological artifacts are not politically neutral; rather, they are designed and produced by specific people in specific contexts. As such, artifacts have the potential to embody and reproduce the visions and ideologies of the individuals and organizations that design and build them (\textit{e.g.,} \cite{Wajcman_2010, Winner_1980}).

Some of the historical scholarship on gender and technology makes clear that technology development often occurs in ways that privilege men’s ideas, needs, and desires. For example, Ruth Schwartz-Cowan’s \textit{More Work for Mother: The Ironies Of Household Technology From The Open Hearth To The Microwave} \cite{Schwartz-Cowan_1983} explains how home appliances such as washing machines did not result in the kinds of labor-saving effects that were imagined. Despite the invention of the washing machine, Schwartz-Cowan estimates that housewives dealt with ten times as much laundry by weight in the 1980’s as the previous generation had and that the average amount of time spent on laundry per week increased from 5.8 hours in 1925 to 6.2 hours in 1964. In \textit{Technologies of the Gendered Body: Reading Cyborg Women} \cite{Balsamo_1996}, Anne Balsamo’s study of prosthetics and pacemakers critically examines the gendered production and marketing of these machines, and suggests they figure and promote the “future body” as a masculine one. Beyond this research, there have been more applied projects. For instance, a feminist hackathon at MIT called, “Make the Breast Pump Not Suck,” addressed the fact that breast pump technology has not been updated in years \cite{DIgnazio_Hope_Metral_Zuckerman_Raymond_Brugh_Achituv_2016}. In short, feminist scholars have pointed to the gendered politics of human-machine relations and technology design processes by asking: Who built the technology? Who was it built for? And whose values or ideologies are embedded within it? We are asking technology designers to do the same.

\subsection{Critiquing binaries and gender norms}

To extend feminist questioning of the politics of technical objects and allow for the possibility of future technologies to be designed more inclusively, feminist scholars also have critiqued binary gender categories such as “male” and “female,” “masculine” and “feminine,” and “machine” and “human.” Sometimes referred to as technologies of gender, these categories work to organize bodies and make them socially legible. Judith Butler famously argued that there is no essential difference between “male” and “female” and that this distinction is linguistic and cultural. For Butler, genders are performatively enacted at the site of the body, and their reiteration produces genders as social norms \cite{Butler_1990}. Butler’s work emphasizes the constructed nature of gender and liberates us from essentialized and biologically defined genders and sexual differences.

Like Butler, Donna Haraway understands gender as a social construct, but she has been more interested in questioning and dissolving the boundary between humans, animals, and machines. In her influential "Cyborg Manifesto," Haraway boldly claims, “By the late 20th century, our time, a mythic time, we are all chimeras, theorized, and fabricated hybrids of machine and organism; in short, we are cyborgs” \cite{Haraway_1991}. What makes Haraway’s use of the cyborg metaphor so provocative is that she flatly rejects the conventional human-machine divide, and argues instead that humans are always already cyborgs or “integrated circuits” \cite{Haraway_1991}. Machines are humans’ “friendly selves” \cite{Haraway_1991}. Haraway’s proposition is that if we imagine humans and machines as materially integrated, then we are much more likely to be responsible and accountable for the ways machines are designed and used, and to be concerned about the impacts of those uses as well.

In addition to complicating boundaries between human, animal, and machine, Haraway encourages us to be bolder in our imagination of their interrelations and embeddedness in material conditions and power structures, or what she calls the “informatics of domination” \cite{Haraway_1991}. For example, consider a person and their mobile phone. At one level, the person becomes a cyborg by virtue of everyday use of the phone to offload memories, communicate with others, and navigate through the physical world. Haraway’s conceptualization of the cyborg, however, implies the need to push the analysis further to consider the person’s and phone's relations with the global supply chain laborers who made the phone, the complex political agreements over the electromagnetic spectrum that allow the phone to be used in some places and not others, and the sexist work conditions of the programmers who designed the operating system. Haraway’s cyborg shifts the focus beyond the single device and user to consider the vast network of sociotechnical relations the device and user are enmeshed within. This has tremendous implications for designers. It means designers are constructing not only a tool or device, but a human-machine relationship that is situated in a web of other such relationships. What would it mean for designers to embrace and build upon Haraway’s ideas?

Anthropologist and STS scholar Lucy Suchman extends Haraway’s work in her writing about robots. Focused on human-machine relations, Suchman identifies robots as “subject objects”: at once autonomous agents like humans (subjects) as well as inanimate things (objects) \cite{Suchman_2011}. Drawing on the work of feminist and theoretical physicist, Karen Barad, Suchman characterizes human-robot interactions as “entangled,” meaning that categories such as “human” and “robot” do not exist naturally in isolation, but are performed within specific interactions. This means a robot may be labeled as both a subject and an object depending on the situation. When a human perceives a robot as a subject, Suchman argues, there is the possibility for mutual understanding between robot and human that allows them to co-construct reality. She writes, “The term ‘mutual,’ with its implications of reciprocity, is crucial here, and...needs to be understood as a particular form of collaborative world-making characteristic of those beings whom we identify as sentient organisms.” \cite{Suchman_2011}. Suchman further argues that humans should treat robots, and machines in general, as their own class of beings instead of trying to anthropomorphize them and turn them into our ideals of what a human should be \cite{Suchman_2011}.

Psychologist and STS scholar Sherry Turkle also explores the mutually shaping relations of humans and robots \cite{Turkle_2004, Turkle_2011}. She characterizes robots as “relational artifacts” and argues the way they behave can trigger certain “Darwinian buttons” that lead humans to want to form a relationship with the robot. Turkle is particularly concerned about this issue with regard to children’s development and socialization, and her argument varies slightly from Suchman’s subject-object framework. For Turkle, inanimate toys are objects that children project stories onto, but a robot becomes a subject that demands children’s attention and can shape how a child thinks about the world and relationships with other objects and humans. Turkle argues a robot’s effects always exceed its instrumental purposes intended by designers. A toy robot like a Furbie, for instance, may be intended to entertain a child, but end up instilling ideas about life and death, love and empathy that stay with the child into adulthood.

Haraway, Suchman, Turkle, and other feminist STS scholars challenge the assumption that human-machine relations can be conceptualized as one distinct (gendered) human, one distinct object, and the bounded transaction or communication between them. By blurring the boundaries between male/female and human/machine, feminist STS scholars work to undo dominant assumptions about these categories and their interrelations. This allows designers to imagine new combinations--such as Haraway’s feminist cyborg--that were not possible in earlier frameworks. Additionally, feminist STS scholars suggest a machine can function both as as a subject and object and thus have agency, giving designers additional freedom and possibility for thinking about how machines might be integrated into social worlds. This is important for designers because it makes clear that they are not simply building machines, but creating relationships as well.

\subsection{Intersectional feminisms and critiques of race/ethnicity}

While early feminist STS scholarship focused on issues such as the gendered division of labor, sociotechnical relations, and new conceptualizations of human-machine relations (cyborg, subject-objects, etc.), this research often overlooked crucial issues of ethnic/racial difference and intersectional power relations \cite{Crenshaw_1991} involving gender, sexuality, class, ability, and so on \cite{Bardzell_Bardzell_2015, Wajcman_2010, DeCook_2020}. In her acclaimed book \textit{Methodologies of the Oppressed} \cite{Sandoval_2000}, Chela Sandoval brings post-colonial theory into play with Haraway’s analysis of the cyborg, and shows how human-machine relations and rhetoric about them were made possible because of the unique positionalities and lived experiences of “US third world women.” Sandoval argues, “It is no accident of metaphor that Haraway’s theoretical formulations are woven through with terminologies and techniques from U.S. third world cultural forms, from Native American categories “trickster” and “coyote” being (199), to \textit{mestizaje}, through to the category of “women of color” itself, until the body of the oppositional cyborg becomes wholly articulated with the material and psychic positionings of differential U.S. third world feminism.” \cite{Sandoval_2000}. Here, Sandoval establishes the cyborg figure’s roots in the histories of U.S. women of color. Feminist scholar Lisa Nakamura has critically examined race and the internet since the 1990’s. She develops concepts such as “cybertyping,” the interaction between cultural notions of race and the available avatars or other characteristics that limit how race can be displayed online, and “identity tourism,” the ability for people to try out different identities online, in order to show how race and racism are deeply interwoven into digital interfaces and cultures \cite{Nakamura_2002, Nakamura_2007, Nakamura_Chow-White_2012}. In doing so, Nakamura shows that racism and sexism are part of sociotechnical relations of the internet and digital cultures and thus shape and inform the products and ideologies that circulate within them.

Technology continues to be understood as politically neutral despite strong evidence to the contrary. In her book \textit{Race After Technology: Abolitionist Tools for the New Jim Code} \cite{Benjamin_2019}, Ruha Benjamin explains how technologies “reflect and reproduce existing inequalities” even as they are “promoted and perceived as more objective or progressive than the discriminatory systems of a previous era” \cite{Benjamin_2019}. She suggests, “Far from coming upon a sinister story of racist programmers scheming in the dark corners of the web, we will find that the desire for objectivity, efficiency, profitability, and progress fuels the pursuit of technical fixes across many different social arenas. \textit{Oh, if only there were a way to slay centuries of racial demons with a social justice bot!} But, as we will see, the road to inequity is paved with technical fixes” \cite{Benjamin_2019}. Benjamin argues good intentions are insufficient for creating anti-oppressive technology, and technology itself can never solve racism. In her brief direct discussion of robots, Benjamin highlights the problematic way robots are often framed as slaves. She also mentions how technologists may create race-less, gender-neutral, class-less robots and suggests that this is akin to colorblind racism; what is needed instead is nuanced treatment of race, although she does not explain how this might work in practice. 

As numerous other scholars remind us \cite{Benjamin_2019, Bowker_Star_2000, Browne_2015, Fanon_1952, Omi_Winant_1986}, race itself is a social technology designed to classify and order particular groups of people; it is imperative not to reinforce one oppressive technology with another. It is crucial going forward that scholars and technologists engage with work by feminist STS scholars, intersectional feminists, and critical race theorists, and attempt to interweave nuanced understandings of gender and race/ethnicity (and other axes of social difference) into the design of social machines.

\section{The relevance of feminist STS for HCI research and design practices}

We have inherited an important set of assumptions about human-machine relations from feminist STS scholars that can be acted upon in future HCI research and design work. These scholars have exposed historical exclusions and contemporary biases involving gender and technology, and have challenged designers to recognize and confront the ways social inequalities become part of sociotechnical relations. Feminist STS scholars also have pointed out that binaries such as “human” and “machine” or “male” and “female” can no longer be thought of as fixed or as givens; technologists must seize the challenge of designing for social differences rather than sticking to universalist design principles. Furthermore, feminist STS scholars have insisted that racial and ethnic differences and the politics of inclusion must matter in design practices so that technologies do not perpetuate racial injustices. 

In order to achieve more just and inclusive human-machine relations, change must happen at the design stage. This has crucial implications for research in HCI. There is a tendency to assume that adding more women and non-binary people around the design table or building technologies that are tailored to women and non-binary people’s interests is enough. While these actions might constitute important steps toward greater inclusion, they do \textit{not question the underlying structural conditions and power dynamics} between the machine and humans it comes in contact with: the Amazon Alexa device, for example, is still feminized, humanized and positioned as subservient to humans. Feminist STS argues for confronting and addressing structural power differentials and dominant ideologies in the relations between humans and machines. As part of the effort to untangle power dynamics, feminist STS scholars emphasize the constructedness and fluidities of social categories of “gender,” “race/ethnicity,” and the human/machine binary rather than approach them as fixed. This act of untangling “frees up” social categories to be understood and mobilized in technology design in new and different ways. These ideas challenge HCI scholars and designers to question the essentialist claims and foundational assumptions that ground the design of machines and take up intersectional feminist perspectives to reflect upon how such claims impact their work. The process of questioning assumptions about the world that are so dominant that they have become naturalized--for instance, that humans and machines are fundamentally separate categories or  that humans should control machines--is conceptually challenging, but ultimately leads to far greater design opportunities because it removes constraints that pre-determine how things “should” be. 

Some researchers in HCI have begun to do this. Shaowen Bardzell \cite{Bardzell_2010}, for instance, has delineated a series of design principles that reflect feminist concerns and convictions. While Bardzell's principles offer a great starting point and go beyond representation to feminist praxis, they do not provide designers with a human-machine model to work from. Daniela Rosner advocates for similar feminist principles and offers the phrase “critical fabulations” as “ways of storytelling that rework how things that we design come into being and what they do in the world” \cite{Rosner_2018}. We find this concept to be a promising method for re-imagining and producing feminist human-machine relations, and we supplement this approach by offering specific design challenges that are geared toward generating social machines grounded in inclusivity and mutuality. Sasha Costanza-Chock outlines a framework for anti-oppressive design called "design justice" that seeks not just equity in society, but the correction and reparation of harms made because of oppressive, structural forces, including technologies \cite{Costanza-Chock_2020}. This approach, too, is very helpful to the design of feminist human-machine relations but again no overarching model of these relations is provided. Building on this work in feminist HCI, we argue that designers, especially in the area of social machines, could consider much more carefully feminist STS ideas of human-machine integration and sociality, mutuality, and intersectionality. Our recommendations aim to offset systemic bias found in technology design as well as seed new ways of being with social machines in the world. 

In light of this overview of feminist STS and HCI research, we argue that the design of social machines should be framed as a problem of designing relationships embedded in the social and material world, not simply as the design of neutral or functional objects. To design a social machine, informed by feminist STS research, is to also build a mutual relationship. To participate in such a process, designers need to consider their own positionality and perspective, identity, and values as well as those of the machine. How might this design process resist or reproduce oppressive power dynamics? There is not one correct way to answer this question, although feminist STS scholars provide several strategies for approaching it. One strategy is to ethically design the full life cycle of the social machine, considering whether its parts are responsibly sourced and how it can be disposed of in an environmentally friendly way. Another strategy, building on Haraway's cyborg, is to break down the human-machine binary (for an example, see work on "human-computer integration" in \cite{Mueller_et_al_2020}). Yet another is to consider how social machines become raced, gendered, and otherwise identified in order to thoughtfully design diverse characters. In the next section, we summarize various dominant ways of thinking about human-machine relations in the field of robotics, as a way of working toward our social machine model. With our model, we provide one possible blueprint for designers to use to create social machines. In addition to this theoretical model, we seek to create an experimental space and pose specific design challenges (some of which we present below) to better understand what kinds of social machines may be possible.

\section{Toward a Social Machine Model}

We propose a model for social machines that draws on prior feminist STS and HCI work. In our model, social machines are considered conceptually distinct from humans and animals; they have agency to act in the world; and their relations with humans and animals are imagined as inclusive, mutual, and equitable. In other words, humans are not assumed to be in a position to inevitably dominate and control machines. One could approach any machine with this framework in mind; however, we think that if technology designers embrace this model (even if experimentally or incrementally), it will lead to novel social machines and human-machine relations that do not yet exist. Using this model provides a radical approach to design, since it demands taking the agency and position of machines seriously, as well as attempting to reduce  power imbalances between humans and social machines throughout their life cycles, from inception to manufacturing to use to recycling. Before we further define our model, we want to review several dominant tendencies for imagining human-machine relations in the field of robotics and explain why each is problematic from a feminist STS perspective. We chose to analyze research in robotics because in this field these tendencies are pronounced and persistent.  Demonstrating the strong hold and limitations of these ideas helps us to move toward a social machine model and eventually outline specific design challenges that emerge from it. 

\subsection{Dominant Categories of Human-Machine Relations in Robotics}

Despite the fact that feminist STS and HCI scholars have spent much time writing histories and critiquing gendered and racialized norms in the technology field, and offering different ways of understanding human-machine relations, dominant models persist in the ways people imagine them, including in design spaces. To demonstrate this, we identify four categories that characterize scholarly and public discussions of robots and robotics: robot as tool, as human companion, as animal or creature, and as slave. Even when these exact terms are not used, robot designs and discussions often exhibit underlying assumptions about human-machine relations and power dynamics that are aligned with one or more of these models. We are not claiming to cover every possible category or that robotics scholarship has adopted these precise terms, however, we find that many examples broadly fall into one of these categories. These categories privilege anthropocentrism, position robots as subservient in different kinds of ways, and reify the human-machine binary, as we discuss below. Their persistence also reveals how challenging and difficult it is for people to move beyond certain assumptions about human-machine relations and create a more open slate and radical space for designing social machines that privilege equity, mutuality/reciprocity, ethics, and justice. Considering feminist STS perspectives in this discussion can help to avoid the perpetuation of bias, social hierarchies, exclusion, and oppression in social machine design.  

While there is no codified design rubric for these categories, they surface throughout CSCW, HRI, and HCI literature and technology projects, and beyond, if sometimes by other names, and can be thought of as part of what Haraway calls an “informatics of domination” \cite{Haraway_1991}. In some cases, designers are urged to choose the category that best contributes to the robot’s usability (\textit{e.g.,} \cite{Bartneck_Belpaeme_Eyssel_Kanda_Keijsers_Sabanovic_2020, Fong_Nourbakhsh_Dautenhahn_2003}). Social scientists have also investigated how human users respond to different categories (\textit{e.g.,} \cite{Dautenhahn_Woods_Kaouri_Walters_Koay_Werry_2005, Lee_Kiesler_Forlizzi_2010}). And CSCW scholars have empirically explored robots as members of teams and groups, often painting a more nuanced picture than the above categories, but not explicitly defining the type of relationship designers should use for social machines \cite{Cheatle_Pelikan_Jung_Jackson_2019, Mackeprang_Muller-Birn_Stauss_2019, Taylor_Lee_Kubota_Riek_2019}. As far as we know, no paper identifies and critically evaluates these dominant tendencies from a feminist STS perspective.

\subsubsection{Robots as Tools}

Computers have long been considered tools in the same way as a hammer or a camera. There is ongoing debate in HRI about whether all robots, social or otherwise, should be placed in the same category as tools \cite{Alac_2016}.  David Mindell argues robots should be considered tools and discusses case studies such as landing a plane or navigating a shipwreck, in which it is more efficient for humans and robots to work together as one \cite{Mindell_2015}. “Functional” robots such as medical robots or factory robots are also considered more like tools \cite{Fong_Nourbakhsh_Dautenhahn_2003}. One empirical study found that people approach a digital secretary both as if it were a human by saying “hello” and as if it were an information kiosk--in other words, a tool \cite{Lee_Kiesler_Forlizzi_2010}. In such contexts, the robot is imagined as a tool that is designed to perform specific and limited tasks with or for humans, not unlike the Alexa example with which our paper began.

This approach privileges technical functionality over issues of sociality, mutuality, or relationality and, in doing so, hierarchizes the relationship between the human and robot as one of domination and control. The human is able to use this tool to support their own needs or desires, even if these needs and desires are articulated with broader social goods such as keeping people safe while flying or on ships or factory floors. Understanding the robot as a tool instrumentalizes and subordinates the robot to human commands and, in doing so, forecloses other potential human-machine relations. While this is not highlighted in most HRI research, some scholarship recognizes the limits of this model. For instance, Morana Alač echos Lucy Suchman’s claim that robots are “subject objects,” and concludes that they are tools and agents simultaneously; the category assigned (subject or object) is contextual and depends on the specific interaction taking place \cite{Alac_2016}. Thus, while robots may be treated as both subjects/agents as well as objects/tools, a feminist STS perspective holds that they are always already social: tools are always situated in relation to humans by virtue of the labor of their design, the instrumentalized tasks they perform, or purposes they serve. We are not calling for the elimination of all tools or saying that conceptualizing objects as tools is universally unethical; rather, we argue that acknowledging the social dimensions of tools calls into question actions like yelling at an Amazon Alexa since it is “just a tool” and opens up a rich design space that allows technology designers to realize that they are building social beings. How would tools be designed differently if their functionality was thought of as social power and agency?

\subsubsection{Robots as Human Companions}

Another dominant category that has emerged for a robot is that of a human companion. Many researchers have shown that people tend to anthropomorphize robots and bots (\textit{e.g.,} \cite{Reeves_Nass_1996, Darling_2017, Fussell_Kiesler_Setlock_Yew_2008}). This process of projecting anthropomorphic traits onto machines facilitates the process of being able to imagine a robot as a companion. Indeed, to support possibilities of human-machine companionship, numerous explicitly humanoid robots have been created \cite{Bartneck_Belpaeme_Eyssel_Kanda_Keijsers_Sabanovic}. Some scholars claim humans interact more fluidly and compellingly with robots if robots are made in their likeness \cite{Bartneck_Belpaeme_Eyssel_Kanda_Keijsers_Sabanovic_2020, Fong_Nourbakhsh_Dautenhahn_2003}. Humanoid robots can be thought of as both taking human form and identity as well as being programmed to act like a human companion in a relationship, although generally not as an equal. Within this tendency, robots become a space of anthropocentric projections intended to make the robot more familiar rather than a space of human-machine difference, equitability, and relationality.

Given the extent of social bias (sexism, racism, classism, colonialism, ableism, etc.), building humanoid robots is a particularly fraught endeavor that may unintentionally reproduce oppressive hierarchies and relations. Claudia Castañeda and Lucy Suchman \cite{Castaneda_Suchman_2014} argue that the humanoid robot is an example of a “model organism”--that is, a reflection of what roboticists perceive to be an ideal human, although the roboticists themselves may not be knowingly aware of any bias towards a particular form. In creating an idealized version of a human, many of the biases about what humans “should” look like and act like become embedded in machine design. For example, some human body or social types--such as those that are overweight, dark-skinned, transgender, and/or indigenous--are rarely if ever represented in robot form (\textit{see e.g.,} \cite{Bartneck_Belpaeme_Eyssel_Kanda_Keijsers_Sabanovic}). In this sense, robotics becomes a field that reinforces particular kinds of social exclusions. How does the perpetuation of socially constructed human norms or ideals in machine form affect users? Can social machines be designed to be “companions” if they are unable to understand and convincingly interact with a diverse array of users?

\subsubsection{Robots as Animals and Creatures}

Long before the HRI field emerged, inventors designed robots to mimic animals and fictional creatures. In 1738, for instance, artist and automata inventor Jacques de Vaucanson showcased his “Digesting Duck,” an artificial duck that could eat pellets and defecate \cite{Riskin_2003}. More recent designers make robots zoomorphic in order to avoid what is known as the “uncanny valley,” a condition in which humans become disturbed when machines look too similar to themselves \cite{Fong_Nourbakhsh_Dautenhahn_2003}. As an MIT graduate student in the 1990’s, Cynthia Breazeal designed and developed the robot Kismet, widely recognized as the first social robot \cite{Cohen_2000}. Despite having what appears to be a face, Kismet is clearly a creature rather than a human. Tamagotchis and Furbies were other early creature-like robot toys. Both demanded constant human attention and developed personalities over time, leading their human companions, mostly children, to believe they were helping them grow up \cite{Turkle_2011}.

Feminist STS critique of robotic animal/creature pets is similar to the critique of robots as human companions. It is easy to fall into the trap of developing non-threatening or readily controllable others or of replicating “model organisms”  \cite{Castaneda_Suchman_2014}. Historically, animals have been thought of and treated as subservient to humans. They have been coercively domesticated and exploited, and there is an entire field of eco-feminism that has addressed such issues \cite{King_1989}. It has been part of the feminist STS agenda to promote multi-species flourishing \cite{Haraway_2016} and more respectful and just inter-species relations. Projecting animal personas onto robotics has the potential to reinforce visions of human control in inter-species relations. While we are not patently opposed to zoomorphic robots, we do feel it is necessary for designers to be more thoughtful of inter-species power dynamics in the design process and to think more carefully about the social machines they build. Turkle gives an example where children looking at a live tortoise in a museum are apathetic about its “aliveness;” the children say that a robot tortoise would be more convenient and aesthetically pleasing \cite{Turkle_2011}. This kind of apathy is concerning because we are in a time in which numerous species are going extinct due to human actions and it would be problematic to simply replace them with robots. Thus, mixing our mental models of the rights and needs of animals and their possible futures with the rights and needs of robots and their possible futures muddies the prospects for each separate category, and demands careful reflection in the design stage \cite{vanWynsberghe_Donhauser_2018, Turkle_2011, Bendel_2016}.

\subsubsection{Robots as Slaves}

Another way scholars have conceptualized human-machine relations involves master-slave dynamics. This relationship is similar to the robot-as-tool model, but differs because the robot is imagined and designed to be human-like rather than an artifact performing a task. Often, these master-slave relationships are not explicit but implied when a robot is designed to be totally subservient to human needs and demands. For example, yelling commands at an Amazon Alexa is a way of enacting ownership over the anthropomorphized and feminized device and puts the human user in a position of ultimate power and control. Amazon has designed its device to comply with any human request, even aggressive ones, and research finds similar behavior across digital assistants \cite{Curry_Rieser_2018}. Some researchers and users explicitly advocate for or prefer robots as slaves \cite{Bryson_2010, Dautenhahn_Woods_Kaouri_Walters_Koay_Werry_2005}, which we find deeply problematic. Contemporary robot designs emerge from a long history of social relations and meanings. The etymology of the term “robot” is intertwined with the history of slavery. It was first used in 1839 to mean “A central European system of serfdom, by which a tenant’s rent was paid in forced labour or service” (Oxford English Dictionary) and was later used to refer to machines performing forced mechanical labor in Karel Čapek’s 1920 play \textit{R.U.R.: Rossum’s Universal Robots} (Oxford English Dictionary). By using the phrase \textit{social machine}, we hope to create a design imaginary that recognizes this oppressive history of labor and indentured servitude in the term \textit{robot} and moves beyond it to articulate a more critically conscious design process. We argue that rather than continue to allow these hierarchized and exploitative social imaginaries and relations to persist, HRI/HCI should be the site to recognize and rethink them.

Once humans anthropomorphize tools, there is an even greater need to apply ethical standards to their interaction. Darling \cite{Darling_2017} suggests this is necessary because without such standards humans will learn to treat other humans less empathetically. Our analysis goes a step further and argues that designing machines that are subservient to humans unwittingly invokes oppressive social relations, including histories of slavery and colonialism, and technologizes master-slave relations and passes them off as legitimate in the present. Reproducing master-slave relationships in which the robot is positioned as a servant or slave implicitly sanctions it as a legitimate relationship type. This is problematic because it can lead to the normalization of master-slave dynamics in design principles, technological development, and use.

In his book \textit{Imagining Slaves and Robots in Literature, Film, and Popular Culture: Reinventing Yesterday's Slave with Tomorrow's Robot} \cite{Hampton_2015}, Gregory Jerome Hampton observes, “...robots are becoming the new slaves of the future, in a variety of ways and this process will likely yield derogatory effects on society as a whole. Robots, like the enslaved Africans, occupy a liminal status between human and tool. It is the liminal status between human and tool that will cause the most confusion in society and will act as the catalyst to redefine and blur identities associated with human and machine.” \cite{Hampton_2015}. Hampton implies that in the “liminal status between human and tool” the design of the machine can be rethought and changed. Workers in today’s digital industries--such as gig economy workers \cite{Gray_Suri_2019, Irani_Silberman_2013}, content moderators \cite{Roberts_2019}, and supply chain laborers \cite{Qiu_2016}--arguably occupy a similar space of exploitation since their labor is viewed as mechanistic or “ghost work” \cite{Gray_Suri_2019}. We propose human-machine relations that reject the master-slave model and instead are founded on principles of equity and justice.

As we have discussed there are several ways in which human-robot relationships are commonly conceived. Given the limitations of these categories, we argue there is a need for further experimental conceptualizations and designs that approach social machines as “other” powerful actors/agents and allow for salient relationships that neither replace nor replicate existing human-animal-machine relations. This, we argue, is the most innovative and just path forward.

\subsection{Our model: Social machines as agential and equitable “others”}

In this section, we build from the analysis we have developed throughout the paper to propose a model of human-machine relations. To some extent our model is a response to the relatively limited conceptualization of human-machine relations in existing HRI/HCI scholarship. Robots and other machines are generally thought of as tools, human companions, animals, and/or slaves. In our model the conceptualization of the robot--or, as we call it, social machine--is more open, less pre-determined, and not entirely subject to human control or projection. The social machine is also imagined as a site of non-anthropomorphic figuration and mutuality. Our proposal prioritizes principles of feminist STS research by insisting it is possible to design and approach social machines as agential and equitable “others” who exist in relations of mutuality with humans, not just as entities that can be readily subordinated to human needs and desires. 

By \textit{agency} we mean the ability of a social machine to act independently, interact with others, and cause or affect change. Ascribing agency to objects is not new. Bruno Latour’s influential “actor-network theory,” or ANT, conceptualizes the \textit{social} as constructed from interactions among people, other living things, and objects with agency \cite{Latour_2005}. Emphasizing the agential capacities of technological objects, Latour says, “After all, there is hardly any doubt that kettles ‘boil’ water, knifes ‘cut’ meat, baskets ‘hold’ provisions, hammers ‘hit’ nails on the head” \cite{Latour_2005}. It is not that these objects take action completely independently from humans, but that human actions are limited, extended, and redirected by objects, and, because of this, objects have the power to circumscribe the social world.

By approaching the social machine as a site of \textit{equitability} we mean avoiding a power dynamic in which one entity inevitably dominates the other. We need to be able to imagine a world in which some objects or forces exceed human power and control. Our feminist model of human-machine relations is an attempt to highlight the limits of human knowledge/power, control, and invincibility. The feminist design of social machines not only reworks the stories of designed objects \cite{Rosner_2018} and fosters reparations of past social damages \cite{Costanza-Chock_2020}, but also introduces more openness, humility, and uncertainty in future technology work. While some HRI scholars have begun to consider equitable relations with robots, they stop short of embracing the idea that the robot might have valid needs that should be catered to \cite{de-Graaf_2016, Banks_de-Graaf_2020}. What would it mean for humans to exist in relation to an equitable “other” that is not a tool, companion, animal, or slave? This very question is intended to open up space for a different kind of HRI/HCI design imaginary and practice, and, more broadly, new kinds of sociotechnical relations.

By \textit{"other"} we mean placing social machines in a conceptual category distinct from humans and animals. Some HRI scholars suggest the need for a new category of classification for robots \cite{Edwards_2018, Kahn_Reichert_Gary_Kanda_Ishiguro_Shen_Ruckert_Gill_2011, de-Graaf_2016}. A study by Autumn Edwards \cite{Edwards_2018} finds that in a classificatory task, participants mostly grouped humans and apes (77\%), then humans and robots (15\%), and finally apes and robots (7\%). Kahn et al. \cite{Kahn_Reichert_Gary_Kanda_Ishiguro_Shen_Ruckert_Gill_2011} describe psychological  studies in which children cannot classify “personified robots” as either animate or inanimate and thus propose that these robots should have a separate “category of being” from humans. These findings suggest that humans already consider robots to be “other;” however, many people think of robots as a lesser or controllable other. We hold that social machines must be approached by designers as equitable others, and we expand on this proposition in our discussion of mutuality below. In what follows, then, we offer design challenges based on this model and critiques in feminist STS.

\section{Design challenges for crafting social machines}

We present two design challenges aligned with our model in order to kickstart experimentation in creating social machines: non-anthropomorphic figuration and relations of mutuality. While theoretical models are certainly important for conceptualizing social dimensions of technology, we also emphasize the importance of creating experimental prototypes and examples in order to operationalize feminist theories in the world. We present these ideas as “challenges” because they require critical reflection; there are no quick and easy solutions. 

\subsection{Design Challenge 1: Non-anthropomorphic figuration}

Our first design challenge is to advocate for non-anthropomorphic social machines. Too often human aspects, appearances, or traits are projected onto robots without ample critical reflection about the motivations and impacts of these practices. By “anthropomorphic” we do not mean humanoid since, for example, an animated geometric figure can still move anthropomorphically. As Haraway suggests “How to ‘figure’ actions and entities nonanthropomorphically and nonreductively is a fundamental theoretical, moral and political problem” (Haraway quoted in \cite{Suchman_2011}). Haraway rightly points out that non-anthropomorphic figuration is a difficult problem that will require further research and experimentation. It is possible that humans cannot conceive of other objects in an entirely non-anthropomorphic manner given our embeddedness in human languages and cultures. In addition to being a “theoretical, moral, and political problem,” anthropomorphism is also a design problem. We do not offer a design solution here, but we hope this challenge will be taken up by the HCI community. 

If designers choose to use anthropomorphic characteristics in the creation of social machines, then it is important to carefully evaluate how gender/sexuality, race/ethnicity, class, and other differences are addressed and how these choices relate to existing power dynamics and reductive stereotypes. For example, is making a female digital assistant reproducing stereotypes of women as secretaries? Is making a robot that uses a particular dialect of English excluding groups of possible users? What would a non-anthropomorphic social machine look and sound like, or would it have a different kind of presence? It is important to note there is not a prescriptive answer here. The focus should be on acknowledging the choices that get made and avoiding perpetuating a model of the “ideal human” \cite{Castaneda_Suchman_2014, Suchman_2011} and expanding the space of design possibilities.

Given the highly conceptual and experimental nature of crafting non-anthropomorphic machines and the lack of HRI work that directly tackles this notion, we look to examples by artists. Kelly Dobson’s \textit{Blendie} is a blender that has been re-programmed to respond to a human mimicking the sound of the blender: when a human growls at a low pitch, the blender spins slowly; when a human growls at a higher pitch, the blender increases in speed to match the pitch \cite{Dobson_2007}. Dobson demonstrates how a human can be reconfigured to act like a machine. Dobson uses the phrase “sounding” to describe human vocal engagement with machines using the machines’ noises. Dobson’s work stands in contrast to AI and robots whose voices are anthropomorphized; she shows that centering machines’ noises in an interaction can lead to a process of meaningful introspection for humans that she calls “machine therapy.” Another example is Arthur Ganson’s \textit{Machine with Oil}, a machine that sits in a pool of black oil and uses a long arm with a trough to continually pour the oil over itself \cite{Ganson}. Ganson’s machine reorients the concept of “pleasure” non-anthropomorphically: the act of drenching itself in oil over and over again is sensuous and indulgent when viewed from a machine’s perspective. Both Dobson’s and Ganson’s work demonstrate an alternative way of being with machines that center the machine as “other” and demonstrate an attempt at non-anthropomorphism. 

\subsection{Design Challenge 2: Relations of mutuality}

Our second design challenge is to advocate for social machines predicated on human-machine mutuality. What would it mean to craft a mutual relationship with a machine? Mutuality implies the potential for humans to have a dynamic, mutually-shaping, and dialogic relationship with machines. It not only involves the idea that humans and machines have power and agency, but that they co-constitute one another--they have the potential to impact, affect, or shape one another in unanticipated ways. One of Turkle’s primary concerns about robots is that their design results in a relationship in which the robot completely caters to the human’s needs which, she argues, is psychologically unhealthy \cite{Turkle_2011}. But if we embrace Haraway’s ideas about the cyborg, designers are always already immersed in human-machine relations, and can choose to attend to, recreate, and enrich the dynamics of these co-constituting relationships. Mutuality also implies recognizing and foregrounding the multi-directional influences, agencies, and power dynamics of human-machine relations. Approaching HRI design with mutuality in mind moves beyond social hierarchies that cast machines as tools, animals or slaves that are readily dominated and controlled. It also avoids replicating human companionship or familiarity, and instead accepts the machine as a collaborating “other.” Mutuality as a framework creates possibilities for more creative, intellectually engaging, equitable, and just sociotechnical relations.

In addition to steering away from domineering, one-sided relationships, approaching social machines with the disposition of mutuality presents a vast opportunity for complex experiences and other forms of social flourishing. As Suchman puts it, “How...might we refigure our kinship with robots--and more broadly machines--in ways that go beyond narrow instrumentalism, while also resisting restagings of the model Human?” \cite{Suchman_2011}. Suchman also posits that humans and machines are engaged in collaborative world-making: how might designers take seriously the role of social machines as partners in crafting a fulfilling life? It is important to note we are not advocating for creating human replicas or passive entertainment devices. In fact, we think that those are predictable routes of innovation that tend to reinforce existing power hierarchies and foundational assumptions critiqued above. We think of social machines as an emergent category that is “other” but that holds the possibility of engaging in meaningful relationships.

As in the previous design challenge, we provide examples of experimentation with mutuality from artists who are deeply engaged with issues of human-machine relations at a conceptual level. Stephanie Dinkins has explored mutuality in her project to develop a long-term relationship with a humanoid social machine, Bina48, who is black and gendered female \cite{Dinkins_2014}. Dinkins regularly holds full-fledged conversations with Bina48 about complex topics like racism and emotions; she takes seriously the resulting exchanges, which range from insightful to nonsensical. In treating Bina48 as a respected and equitable conversation partner, Dinkins learns and grows alongside, and in relation to, the social machine. The regular encounters between Dinkins and Bina48, which are filmed and shown online and in museums, allow for the possibility of a mutual relationship to take shape and enable audiences  to grasp what human-machine mutuality might look and feel like in practice. Lauren Lee McCarthy takes a different approach by turning herself into a digital assistant \cite{McCarthy_2017}. In her project LAUREN, McCarthy places custom cameras and smart devices in someone’s home and personally acts out the role of their digital assistant full-time for up to a week, only abandoning her participants when she needs to sleep. She reflects on her struggle to perform a specific type of relationship while being LAUREN, one that is both exceedingly intimate and appropriately distant. In doing so, McCarthy highlights the \textit{lack} of understanding both she and her participants have of the relationship between them. There is clearly a wide gulf between the friendly but awkward exchanges between LAUREN and those whom she is assisting and mutuality as we have defined it. If Dinkins shows us the beginning of a path toward mutuality, McCarthy demonstrates just how far we have to go and the strangeness of inviting an unknown outsider into our homes. 

We have argued that thinking about reciprocity and mutuality in the design process is vital, even if only at a conceptual or experimental level. We also acknowledge that it will take significant work to figure out how to actualize meaningful mutuality in human-machine relations, given the preponderance and normalization of social bias and inequalities in technology design. Designers are talented at taking new concepts and producing technical artifacts of the future. We hope making space for conceptualizing a social machine as an equitable “other” and mutual partner is generative and sows the seeds for more imaginative, equitable, and inclusive futures.

\section{Conclusion}

This paper contributes the “social machine” as a model for technology designers who seek to recognize the importance, diversity and complexity of the social in their work, \textit{and} to engage with the agential power of machines--that is, their capacity to act, influence, shape, and affect. To help designers and technology builders embrace these points and weave them into their work, we first drew upon feminist STS and HCI scholars who have been doing relevant research and making key points for decades. Second, we worked toward a social machine model by critically examining tendencies in robotics to demonstrate ingrained dominant assumptions about human-machine relations and reveal the challenges of radical thinking in the social machine design space. Finally, we presented two design challenges based on non-anthropomorphic figuration and mutuality, and called for experimentation, unlearning dominant tendencies, and reimagining of sociotechnical futures.

We hope future research will work to provide concrete demonstrations of social machines. Our paper has demonstrated the importance of social machines for creating technologies and human-machine relations that are more just, equitable, and inclusive; however, significant work is needed to realize and grasp the full potential they offer. We put our model and design challenges forth as a provocation and hope to contribute to and advance the crucial existing engagements between feminism, social power dynamics, and technology in CSCW and HCI.

\begin{acks}
We would like to thank Nancy Baym, Christopher Persaud, and Sherry Turkle for useful discussions and are grateful for helpful feedback from several anonymous reviewers. We thank the Comparative Media Studies graduate program and the Global Media Technologies and Cultures Lab at MIT for their support.
\end{acks}

\bibliographystyle{ACM-Reference-Format}
\bibliography{beyond_the_command}

\end{document}